\documentclass{mpe_report}

\usepackage{psfig,graphicx,epsfig}
\usepackage{color}
\usepackage{amsmath,amssymb,epic,eepic,array}

\begin{document}

\pagenumbering{arabic}
\setcounter{page}{28}

\renewcommand{\FirstPageOfPaper }{ 28}\renewcommand{\LastPageOfPaper }{ 31}


\title{Upper Limits on the Pulsed VHE $\gamma$-ray Emission from Two Young Pulsars Investigated with the High Energy Stereoscopic System}
\author{A. Noutsos\inst{1,}\inst{2} for the H.E.S.S.~collaboration}  
\institute{Jodrell Bank, Macclesfield, Cheshire SK11 9DL, UK
\and  Department of Physics, Durham University, South Road, Durham DH1 3LE, UK}
\maketitle

\begin{abstract}
The High Energy Stereoscopic System (H.E.S.S.) is a system of four, imaging, atmospheric Cherenkov telescopes in Namibia, designed to detect very-high-energy $\gamma$ rays above $\sim 100$ GeV. During 2002--2003, H.E.S.S.~collected data from two, young and energetic radio pulsars: the Crab and PSR B1706$-$44. We searched for pulsations at the lowest energies that H.E.S.S.~is capable of detecting, aiming at a detection that would potentially differentiate between the two popular models of pulsar high-energy emission: the Polar Cap and the Outer Gap. No evidence for pulsed emission was found in the data, and upper limits were derived to a 99.95\% confidence level. Our assumptions and upper limit values for the two pulsars are reported.
\end{abstract}

\section{Introduction}
The Crab pulsar (PSR B0531+21) is a well-known source of electromagnetic pulsed emission in a broad range of frequencies, from radio to high-energy (HE $\gtrsim$ 100 MeV) $\gamma$ rays (Thompson 1999). Virtually all efforts to detect this emission has led to success, and this pulsar has been justifiably characterized as the ``standard candle''. EGRET has detected its pulsed signature up to $\sim$ 20 GeV (Nolan et al. 1993). Yet, Crab's pulses become undetectable above this range, and so far there has not been a verifiable detection in very high energies (VHE $\gtrsim$ 100 GeV).

PSR B1706$-$44 is a young, southern-hemisphere pulsar that lies relatively close to the Galactic centre. It was first discovered in radio, but it also appears as a strong emitter in HE $\gamma$ rays (Johnston et al. 1992; Thompson et al. 1996). The detection with EGRET in the latter energy range has provided the only confident high-energy profile so far, although a strong indication of pulsed emission, to a 4$\sigma$ level, was also found in X-ray data from {\em Chandra} (Gotthelf et al. 2002). Above $\sim$ 20 GeV, only upper limits exist for this pulsar.

A few of the above pulsars' properties are summarised in Table~\ref{tab:properties}.

\section{Theory}
The absence of pulsed emission above EGRET's range is consistent with two popular classes of theoretical models, the Polar Cap (PC) and the Outer Gap (OG). They predict that the high-energy spectra of pulsars should diminish rapidly beyond a certain cut-off energy.

\subsection{Polar Cap models}
\begin{figure}
\centerline{\psfig{file=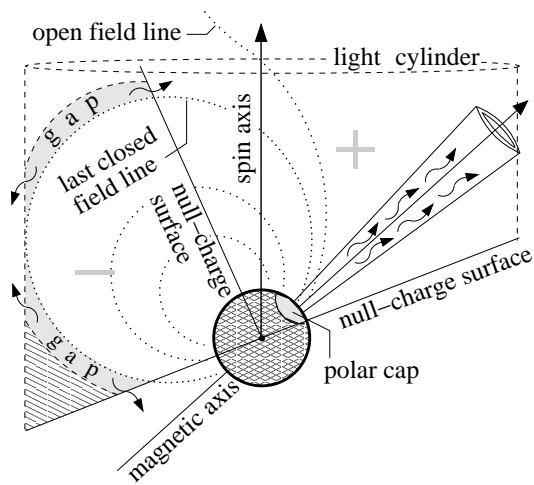,width=7cm,clip=} }
\caption{Geometrical model of a pulsar's magnetosphere with the locations of emission according to the Polar Cap and Outer Gap models.
\label{fig:pulsarGeom}}
\end{figure}

According to PC models, the observed pulsed emission is generated close to the pulsar's surface, above the polar caps (Fig.~\ref{fig:pulsarGeom}). High-energy $\gamma$-ray production is triggered mainly via synchro-curvature radiation from relativistic e$^{-}$ and e$^{+}$. Such models predict HE spectra with steep cut-offs at a few GeV (Sturrock 1971; Ruderman \& Sutherland 1975).

\subsection{Outer Gap models}
The OG models, on the other hand, place the site of emission in the outer magnetosphere, inside charge-depleted regions (gaps) near the null-charge surface (Fig.~\ref{fig:pulsarGeom}). The main radiation mechanism is synchro-curvature radiation from e$^{-}$--e$^{+}$ pairs accelerated along the magnetic field lines inside the gaps. Furthermore, OG models predict an additional inverse Compton component from the upscattering of low-energy photons (i.e.~IR, optical, X-ray) by accelerated e$^{-}$--e$^{+}$ pairs. In general, OG models produce gentler spectral cut-offs than those of PC models (Cheng, Ho \& Ruderman 1986; Chiang \& Romani 1992).

\begin{table}
      \caption{The properties of the Crab and PSR B1706$-$44 pulsars}
         \label{tab:properties}
      \[
         \begin{array}{p{0.5\linewidth}rr}
            \hline
            \noalign{\smallskip}
                                                          &   {\bf Crab}  &{\bf PSR \ B1706}{\rm -}{\bf 44}\\
            \noalign{\smallskip}
            \hline
            \noalign{\smallskip}
             $P$ (s)                                     & 0.033                  & 0.102                \\
             $\dot{P}$ ($\times$ 10$^{-15}$ s s$^{-1}$)  & 421    & 93        \\ 
             $\log{\tau}$ (y)                  & 3.1^{\rm a}                   & 4.2     \\
             $\log B_{\rm s}$ (G)                        & 12.6                 & 12.5              \\
             $\dot{E}$ ($\times$ 10$^{36}$ erg s$^{-1}$) & 450                   & 3.4                 \\
             $\epsilon_{\gamma}$\% ($>$ 100 MeV)$^{\rm b}$          & \sim 0.01                 & \sim 0.1   \\
            \noalign{\smallskip}
             \hline
         \end{array}
      \]
\begin{list}{}{}
\item[$^{\rm a}$] This is the dynamic age, $\frac{1}{2}(P/\dot{P})$; Crab's true age is 952 y. 
\item[$^{\rm b}$] HE $\gamma$-ray conversion efficiency: $L_{\gamma}(>100 \ {\rm MeV})/\dot{E}$
\end{list}
\end{table}

\section{Observations}
H.E.S.S. is a system of four imaging Cherenkov telescopes at the Khomas highland of Namibia (Aharonian et al. 1997b). The array is capable of detecting, at 5$\sigma$ significance, TeV point sources that are 100-times weaker than the Crab nebula, in 50 h exposure time.  At its threshold ($\sim$ 100 GeV), H.E.S.S. can still detect sources that are 10-times weaker than the Crab nebula, given the same exposure time.

In 2003, H.E.S.S. performed {\em stereoscopic} observations of the Crab, using three of the four telescopes of the array. During that time, the source was observable between 45$^\circ$ and 60$^\circ$ zenith angle ($Z.A.$). For our analysis, we selected only data fulfilling quality criteria, like good weather conditions and high trigger rate ($R>150$ Hz). The total exposure time of the selected data set amounted to $T\approx$ 4.5 h.

In 2002, H.E.S.S. performed {\em single-telescope} observations of PSR B1706$-$44, using the first telescope available then. The $Z.A.$ of the observations was $<30^\circ$. Based on the same quality criteria, an extensive data set comprising 28 h of data was selected for analysis.   
 
\section{Data analysis}
\subsection{Low-energy cuts}
Our analysis targeted the lowest energies that H.E.S.S. could detect in the above observations, in an effort to intercept the tail-end of the pulsed emission observed with EGRET. For that purpose, we applied specially tailored event-selection cuts to the $\gamma$-ray images: we rejected all events whose total photo-electron content was above 100 and accepted only events which laid closer than 18 mrad from the source position. After the application of cuts, the background level was $\sim 10^6$ events for both data sets.

\subsection{Temporal analysis}
\begin{figure}
\centerline{\psfig{file=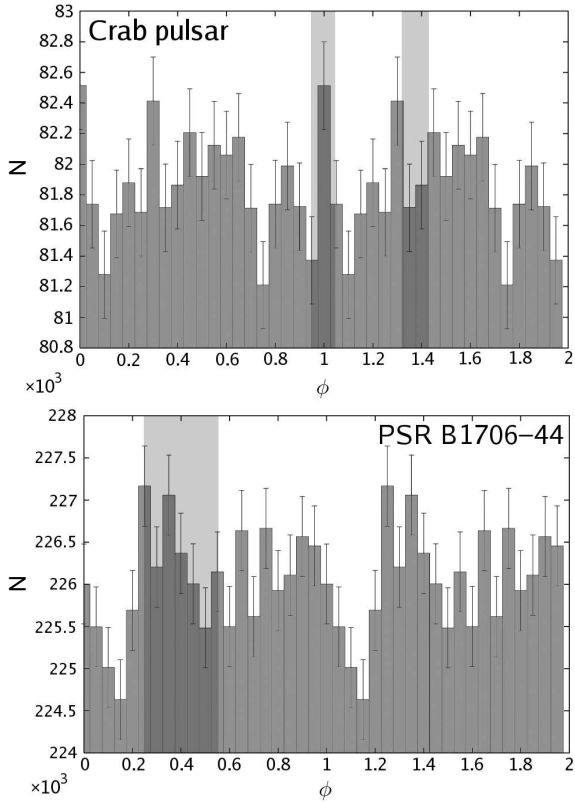,width=7.8cm,clip=} }
\caption{Phasograms of the folded events from 4 h of Crab pulsar data and 28 h of PSR B1706$-$44 data. The shaded areas correspond to the phase ranges where significant pulsed emission was seen with EGRET. 
\label{fig:phasograms}}
\end{figure}

Using standard epoch-folding techniques (EF), the times of arrival (TOAs) of the events were folded into phases. For that purpose, an accurate, contemporaneous ephemeris was needed: for the Crab, it was obtained from Jodrell Bank, and for PSR B1706$-$44 we used the GRO ephemeris of the Australian Pulsar Timing Archive. The resulting phasograms after binning the event phases across two pulsar periods are shown in Fig.~\ref{fig:phasograms}. Both phasograms appear `flat' and consistent, within the errors shown, with Poissonian-background fluctuations. 

Despite the apparent uniformity, we decided to search for a potential signal, hidden in the large background, with more sensitive tests for uniformity: those included e.g.~the $H$-test and the $\chi^2$-test (de Jager et al. 1989). The chance probabilities from the $H$-test are shown in Table~\ref{tab:results}. On the whole, all chance probabilities from all our tests were consistent with the ``null hypothesis'': i.e. the lack of periodicity in the data. 

\subsection{Upper limits on the pulsed flux}
The uniformity tests did not reveal significant deviations from the background. However, this could be due to the relatively weak signal compared to the background fluctuations. Hence, knowing the magnitude of the latter, one can set upper limits (hereafter ULs) to the value of the pulsed flux. 

We decided to calculate 3$\sigma$ (or 99.95\%) ULs on the $\gamma$-ray counts, using the method of O.~Helene (1983). Helene's method requires prior knowledge of the phase ranges where pulsed emission occurs (``peak areas''), across the pulsar's period. For the Crab and PSR B1706$-$44, the peak areas are known from HE observations with EGRET (see Fig.~\ref{fig:phasograms}); we assumed that those ranges remain the same in the VHE regime. 

In order to translate the above ULs to flux ULs, we utilised Monte Carlo simulations to estimate the telescope's effective area ($A_{\rm eff}$) for those events passing our cuts. The $A_{\rm eff}(E)$ functions were derived for $Z.A.$ corresponding to the observed data. 

\subsection{Energy thresholds}
An important parameter that would help quantify how low in energy our cuts have allowed us to probe is the energy threshold of the data sets after cuts. Typically, the energy threshold for a particular sample of events --- with a certain range of energies, incident angles, distances from the telescope, etc. --- represents the energy below which the detector's $A_{\rm eff}$ for these events declines rapidly. More accurately, the energy threshold is defined as the maximum of the {\em differential trigger rate:} the latter being equal to the product between $A_{\rm eff}$ and the differential flux spectrum, $dN/dE$. For the latter, we simply assumed an extrapolation of EGRET's HE power laws, $dN/dE \propto E^{-\nu}$, beyond $\sim 20$ GeV. The assumed spectral indices, $\nu$, are shown in Table~\ref{tab:results}. 

\begin{figure}
\centerline{\psfig{file=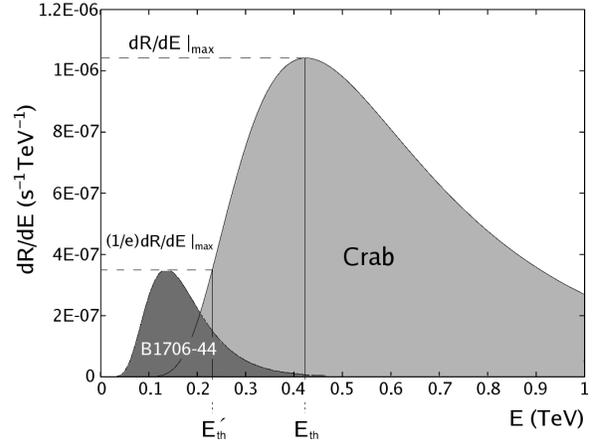,width=7.8cm,clip=} }
\caption{Differential trigger rate functions ($dR/dE$) for the Crab and PSR B1706$-$44, as a function of photon energy, for the assumed power-law spectra with $\nu=2.08$ and $\nu=2.1$, respectively. Also shown are the standard definition of the energy threshold ($E_{\rm th}$), as the maximum of $dR/dE$, and our definition  of $E^{\prime}_{\rm th}$, at 1/e of the maximum $dR/dE$.
\label{fig:diffRates}}
\end{figure}

By convolving $A_{\rm eff}$ with $dN/dE$, we plotted the differential trigger rates, $dR/dE$, for the two pulsars (see Fig.~\ref{fig:diffRates}). It can be seen that, although there is a clear maximum at an energy $E_{\rm th}$ in both cases, the low-energy tail below $E_{\rm th}$ accounts for a large fraction of the total trigger rate; and, clearly, this part must be considered in the calculation of our ULs. Hence, we decided to use a more representative definition of $E_{\rm th}$ in our analysis: we set $E^{\prime}_{\rm th}$ to be the energy at which $dR/dE$ drops to 1/e of the maximum value, as one moves down in energy. This led to the thresholds shown in Table~\ref{tab:results}.   

\begin{table}
      \caption{Analysis parameters and resulting fluxes for the Crab and PSR B1706$-$44}
         \label{tab:results}
      \[
         \begin{array}{p{0.5\linewidth}rr}
            \hline
            \noalign{\smallskip}
                                                          &   {\bf Crab}  &{\bf PSR \ B1706}{\rm -}{\bf 44}\\
            \noalign{\smallskip}
            \hline
            \noalign{\smallskip}
             $T$ (h)                                      & 4.5                  & 28                \\
             $Z.A.$                                         & 45{\rm -}60^\circ    & <30^\circ        \\ 
             $R_{\rm background}$ (Hz)$^{\rm a}$     & 98                   & 45     \\
             $p^{\rm b}$                                          & 0.29                 & 0.008              \\
             $\delta \ (\% P)^{\rm c}$                              & 10                   & 30                 \\
             $\nu$                                        & 2.08                 & 2.1                \\
             $E^{\prime}_{\rm th}$ (GeV)                  & 232\pm 51            & 75\pm 12     \\
             $A_{\rm eff}(E^{\prime}_{\rm th})$ (m$^2$)   & 10^4                 & 400                \\
             $F_{\rm ul}(>E^{\prime}_{\rm th})$ ($\times$ 10$^{-10}$ cm$^{-2}$ s$^{-1}$) & 1.8 & 25  \\
            \noalign{\smallskip}
             \hline
         \end{array}
      \]
\begin{list}{}{}
\item[$^{\rm a}$] Background trigger rate after cuts 
\item[$^{\rm b}$] Chance probability under the $H$-test  
\item[$^{\rm c}$] Duty cycle of the pulsed emission observed with EGRET above 100 MeV

\end{list}
\end{table} 

\begin{figure}
\centerline{\psfig{file=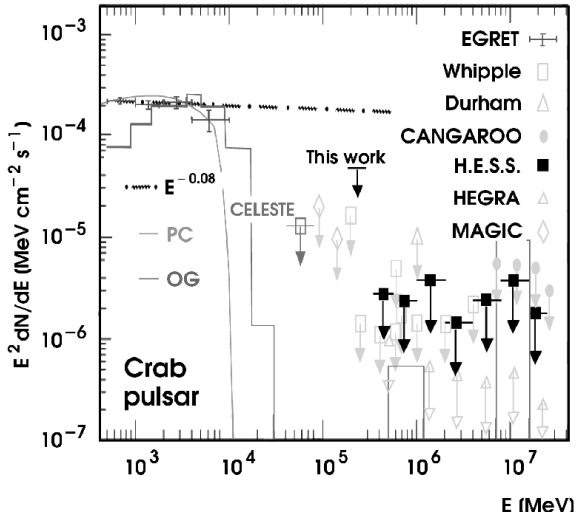,width=7.8cm,clip=} }
\caption{The derived upper limit (UL) on the differential pulsed $\gamma$-ray flux of the Crab pulsar, after the application of our low-energy cuts (``This work''). Flux ULs from various experiments are shown in light grey, with different symbols (Lessard et al. 2000; Dowthwaite et al. 1984; Tanimori et al. 1998; Aharonian et al. 2004; L\'opez 2005). The UL at 60 GeV has been derived from observations with CELESTE (de Naurois et al. 2002). All ULs labelled ``H.E.S.S.'' have been derived from the {\em standard} H.E.S.S.~analysis (solid black squares) (Schmidt 2005). Also, the Polar Cap (PC) and Outer Gap (OG) model spectra for this pulsar are plotted with a smooth, light-grey and a stepped, grey line, respectively (Schmidt 2005; Hirotani \& Shibata 2001). The squiggled--dotted line is a power-law fit to EGRET's data points, extrapolated to very high energies (Nolan et al. 1993).    	
\label{fig:CrabUplims}}
\end{figure}

\begin{figure}
\centerline{\psfig{file=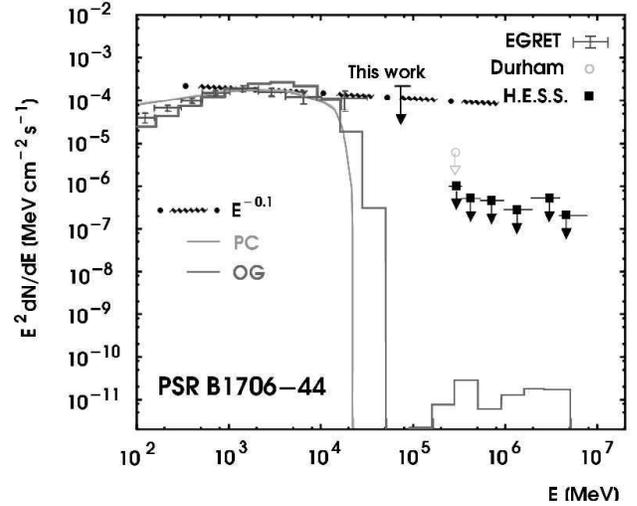,width=8.2cm,clip=} }
\caption{The derived upper limit (UL) on the differential pulsed $\gamma$-ray flux of PSR B1706$-$44, after the application of our low-energy cuts (``This work''). The UL from observations with Durham's Mark 6 telescope is shown with a grey, empty circle (Chadwick et al. 1998). All ULs labelled ``H.E.S.S.'' have been derived from the {\em standard} H.E.S.S.~analysis (solid black squares) (Schmidt 2005). Also, the Polar Cap (PC) and Outer Gap (OG) model spectra for this pulsar are plotted with a smooth, light-grey and a stepped, grey line, respectively (Schmidt 2005; Hirotani 2003). The squiggled--dotted line is a power-law fit to EGRET's data points, extrapolated to very high energies (Thompson et al. 1996).  	
\label{fig:1706Uplims}}
\end{figure}

\section{Results and conclusions}
Our ULs for the Crab and PSR B1706$-$44 are shown in Fig.~\ref{fig:CrabUplims} and \ref{fig:1706Uplims}. Up to $\sim$ 20 GeV, there exist confident detections with EGRET, whereas at higher energies the emission has only been constrained with ULs from various VHE ground-based experiments. It can be seen that our ULs are well-below the energies of those after the {\em standard} H.E.S.S.~analysis and below many of other experiments. Unfortunately, due to the large background inherent in our low-energy analysis, the corresponding ULs are 2--3 orders of magnitude larger than those at higher energies.  

In the case of the Crab pulsar, we can conclude that the derived UL excludes confidently the possibility of a single power law from EGRET's range to at least $\sim$ 230 GeV, thus verifying the indications for a cut-off already observed at the top energy bin of EGRET. However, for PSR B1706$-$44, the large UL value prevents a similar conclusion.

Alongside the experimental results, we have also plotted the spectral predictions from a PC and OG scenario for these two pulsars. One of the main differences between PC and OG spectra is the much steeper cut-offs of the former compared to the latter; this could provide the key to determining which model describes pulsar emission best. However, it is clear that our ULs, although nearer to the spectral cut-offs than previous H.E.S.S.~results, still lack the sensitivity in terms of flux at the required low energies; hence the problem stands.

Future experiments, like H.E.S.S. Phase II, MAGIC and {\em GLAST}, will almost certainly fill the energy gap between HE and VHE observations and, consequently, piece the puzzle of pulsar $\gamma$-ray emission.


          \clearpage

\end{document}